**Anomalous increase in nematic-isotropic transition temperature in dimer molecules induced by magnetic field**

S. M. Salili[a], M. G. Tamba[b], S. N. Sprunt[c], C. Welch[d], G. H. Mehl[d], A. Jákli[a] and J. T. Gleeson[c]

*[a] Chemical Physics Interdisciplinary Program & Liquid Crystal Institute, Kent State University, Kent, OH 44242, USA*

*[b] Department of Nonlinear Phenomena, Institute for Experimental Physics, Otto von Guericke University Magdeburg, Magdeburg, Germany*

*[c] Department of Physics, Kent State University, Kent, OH 44242, USA*

*[d] Department of Chemistry, University of Hull, Hull HU6 7RX, UK*

***Abstract***

*We have determined the nematic-isotropic transition temperature as a function of applied magnetic field in three different thermotropic liquid crystalline dimers. These molecules are comprised of two rigid calamitic moieties joined end to end by flexible spacers with odd numbers of methylene groups. They show an unprecedented magnetic field enhancement of nematic order in that the transition temperature is increased by up to 15K when subjected to 22T magnetic field. The increase is conjectured to be caused by a magnetic field-induced decrease of the average bend angle in the aliphatic spacers connecting the rigid mesogenic units of the dimers.*

Nematic liquid crystals (NLC) are anisotropic fluids that only exhibit uniaxial, apolar orientational order. In most liquid crystals, this order is temperature dependent, spontaneously arising at temperatures below the nematic-isotropic phase transition temperature ($T_{NI}$); above this temperature the material exhibits no order (i.e. is isotropic). In many nematics, orientational order is particularly responsive to external influences, for example electric and/or magnetic fields, mechanical strains, etc. [1–7]. This response is the primary reason that nematic liquid crystals are extensively used in information display applications like liquid crystal displays (LCD's). External fields affect

NLCs in two ways: they affect the degree of orientation order, or they re-orient the axis of orientational order (the “director”, represented by a unit vector field $\hat{n}$ ).

The former effect was demonstrated with electric field by Helfrich [8], who observed an increase in $T_{NI}$ when a large electric field was applied to an NLC. This increase scaled with $E^2$, as was expected (see discussion below); however, it did not exceed 1ºC, even at the largest electric fields used. This result has since been confirmed for numerous other materials. Indeed, mean field theories of the N-I transition predict a critical field at which the transition becomes continuous; this has also been observed for electric fields. [9] Lyotropic liquid crystal materials, in which solute concentration is more important than temperature in determining phase behavior also exhibit field enhanced order. [10].

Magnetic field effects on orientational order are more difficult to observe because the diamagnetic anisotropy is effectively smaller than the dielectric anisotropy. In calamitic (rod-shaped) liquid crystals, applying a 10T magnetic field only increases $T_{NI}$ by a few mK. [11] In these materials, the critical magnetic field is estimated to exceed 100T. More recent work examining NLC’s composed of less linear and reduced symmetry molecules, such as bent-core molecules revealed larger (on the order of 1K) shifts in $T_{NI}$. [12,13]. Neither of these results are explicable within the context of classical mean-field theories for the NI transition, such as Landau-deGennes or Maier-Saupe theories. Instead, the large shift was attributed to additional degrees of freedom: for example, the presence of local polar order or fluctuations of positional order. A detailed calculation of the effects of molecular biaxiality found that this may also lead to anomalously large field effects for the N-I transition. [14]

In this work, we report on an unprecedented magnetic field-induced shifts of the isotropic-nematic phase transition temperature observed in liquid crystal dimers where two rigid linear mesogens are linked by flexible nonyl or heptyl chains. [15] The shapes of these molecules resemble *nunchaku* fighting sticks. The three compounds studied and their corresponding phase sequences are depicted in Figure 1(a). The first compound 1”,7”-bis(4- cyanobiphenyl-4’-yl)nonane (CB9CB), was synthesized as described in Ref. [16]. The synthesis of the second and third dimers, 1,1,1-di(2’,3”-

difluoro-4-pentyl[1,1';4',1"]terphen-1"-yl)nonane (DTC5C9) and 1,1,1-di(2',3"-difluoro-4-pentyl[1,1';4',1"]terphen-1"-yl)heptane (DTC5C7), is described in Ref. [17]. The three compounds exhibit the recently discovered twist-bend nematic ($N_{tb}$) liquid crystal phase, in which the director follows an ambidextrous conical helix with a remarkably short pitch in the 10 nm range. [18] [19] Thus the $N_{tb}$ phase can be considered as a nanoscale pseudolayered structure [20–22], which explains the smectic like textures [20–24] and viscoelastic properties [25] of the $N_{tb}$ phase, despite the lack of a density modulation characteristic of a true smectic. [18,24,26]

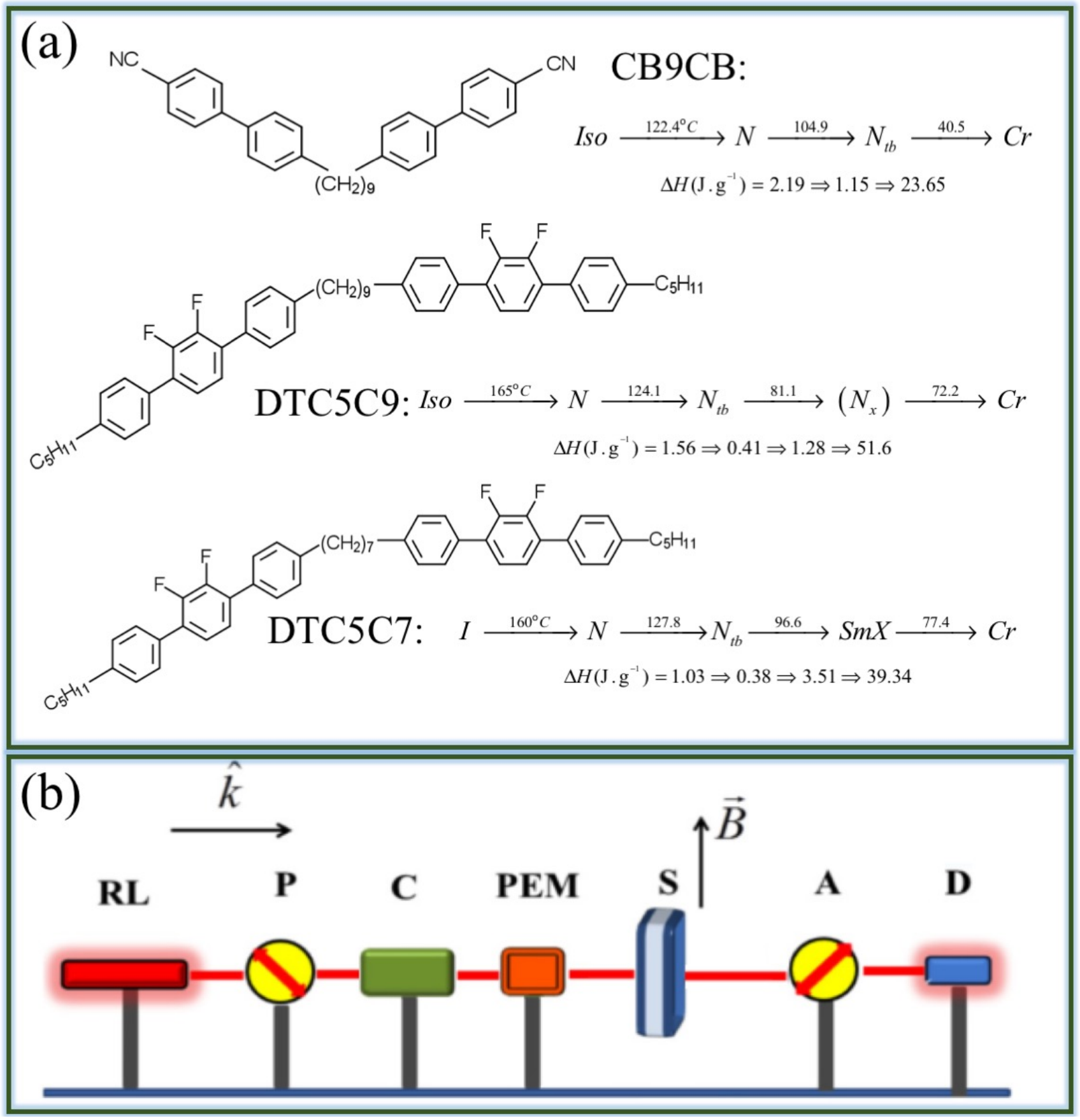

*Figure 1: Studied materials and experimental setup. (a) Chemical structures, abbreviated names and phase sequences of the mesogens investigated. (b) Experimental setup. RL: 30 mW, λ = 632.8 nm He-Ne laser; S: LC sample placed between crossed polarizers (P and A) oriented at ±45° with respect to the vertical magnetic field C: A compensator to correct for residual birefringence; PEM: photoelastic modulator; D: Photodetector positioned to measure the direct transmitted beam. Using this arrangement, the protocol is to either measure the effective birefringence as a function of applied field at fixed temperature, or, the effective birefringence as a function of temperature at fixed field.*

In our experiments we load each compound in a 10 $\times$ 10 mm planar glass cell whose inner surfaces are treated with a unidirectional rubbed polyimide PI2555 (HD Micro Systems) that promotes molecular alignment parallel to the substrates (homogeneous planar alignment) and along the rubbing direction; the distance *d* between substrates was 5 μm. The filled liquid crystal cell is then inserted into a Teflon-insulated temperature-controlled oven (temperature stability $\pm 0.05$ºC) with optical access. Two high-precision temperature sensor probes are embedded in the oven, one is a glass-encapsulated thermistor and the other a platinum RTD. Neither showed any drift during application of high magnetic fields. The oven is inserted into the bore of the 25 Tesla split-helix resistive solenoid magnet at the National High Magnetic Field Laboratory (NHMFL). [27] This magnet has ports allowing optical access perpendicular to the field direction. The oven and the LC cell inside are oriented so that the optical path is orthogonal to the field. The rubbing direction of the planar cell (optic axis) is parallel to the field. A standard optical setup (Figure 1(b)), incorporating photo-elastic modulator and lock-in amplifiers, was used to measure field induced changes in optical birefringence [28]. This setup records the phase difference between ordinary and extraordinary rays, $\varphi = 2\pi \Delta n d / \lambda$, where $\Delta n$ is the effective birefringence of the sample [28–30] and λ is the wavelength of light (632.8 nm).

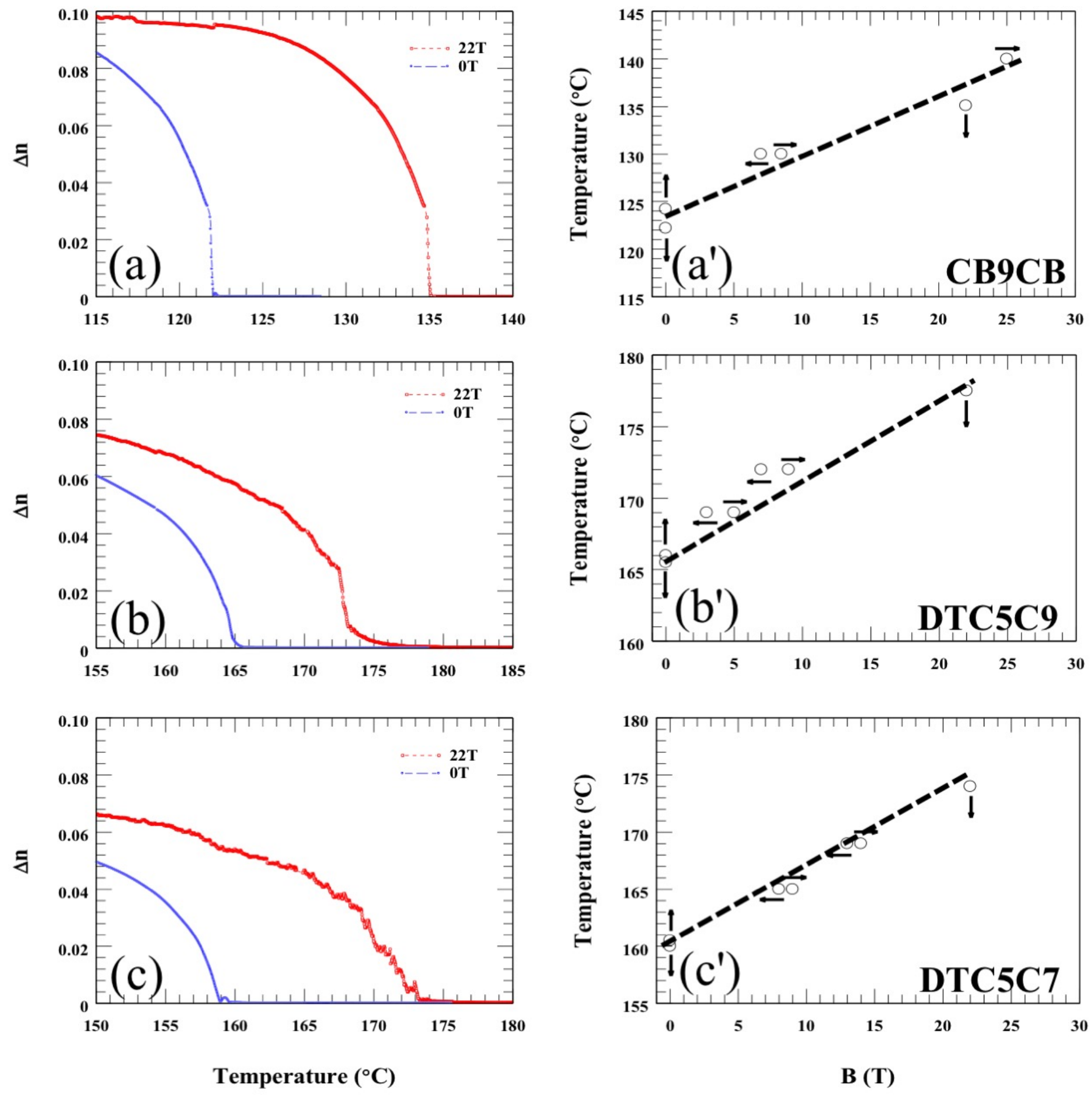

*Figure 2. Temperature dependence of effective birefringence and magnetic field dependence of the I-N transition temperatures. Left: Temperature dependence of birefringence at B=0 (blue) and B=22T (red) for CB9CB (a), DTC5C9 (b) and DTC5C7 (c) measured on cooling at 2 °C/min rate. Right: $T_{I-N}$ as a function of magnetic field B for for CB9CB (a'), DTC5C9 (b') and DTC5C7 (c')***.** *Data with arrows (↑,↓,→ and ← are taken on heating, cooling, increasing fields and decreasing fields, respectively).*

The temperature dependencies (on cooling) of the effective birefringence $\Delta n$ and the dependence of $T_{NI}$ on magnetic fields are shown in Figure 2. At zero field, the discontinuity in birefringence at $T_{NI}$ is plainly visible in *CB9CB*; it is smaller for the two other compounds. The data shown correspond to cooling from above $T_{NI}$. The same behavior is observed in heating, although the transition temperature is observed to be approximately 1ºC higher.

In a 22T magnetic field, $T_{NI}$ increases dramatically, by at least 8ºC in all three compounds, and by almost 15ºC for *DTC5C7*. We know of no other example of such a large magnetic field effect on a nematic-isotropic transition for a thermotropic liquid crystal. Indeed, this effect is more than 300 times larger than what has been reported previously for rod-shaped thermotropic compounds. We also performed measurements at several weaker magnetic fields. As can be seen on the right-hand side of Figure 2, the transition temperature shift $\Delta T_{NI}$ increases almost proportionally to $B$, but is not proportional to $B^2$, as was found for other thermotropic materials. [11], [13]

When the samples are held at temperatures just above the zero field $T_{NI}$ and are then subjected to an increasing magnetic field, the field induces the nematic phase. If the field is subsequently reduced to zero, the isotropic phase returns at a field just slightly lower than that at which the nematic phase was induced. No discontinuity in birefringence (as in Figure 2) is observed in this field-induced transition (see Figure 3), due to the fact that the field was continuously changing (typically at 5T/minute) during the measurements. At temperatures further away from the zero-field $T_{NI}$ the threshold field where the I-N transition is induced increases, as seen in Figure 3 (a-c). In Figure 3(d) we compare $\Delta$n(B) of the three studied materials at 5K above the zero-field $T_{NI}$. The compound CB9CB shows a more rapid onset of birefringence with applied magnetic field compared to the other compounds.

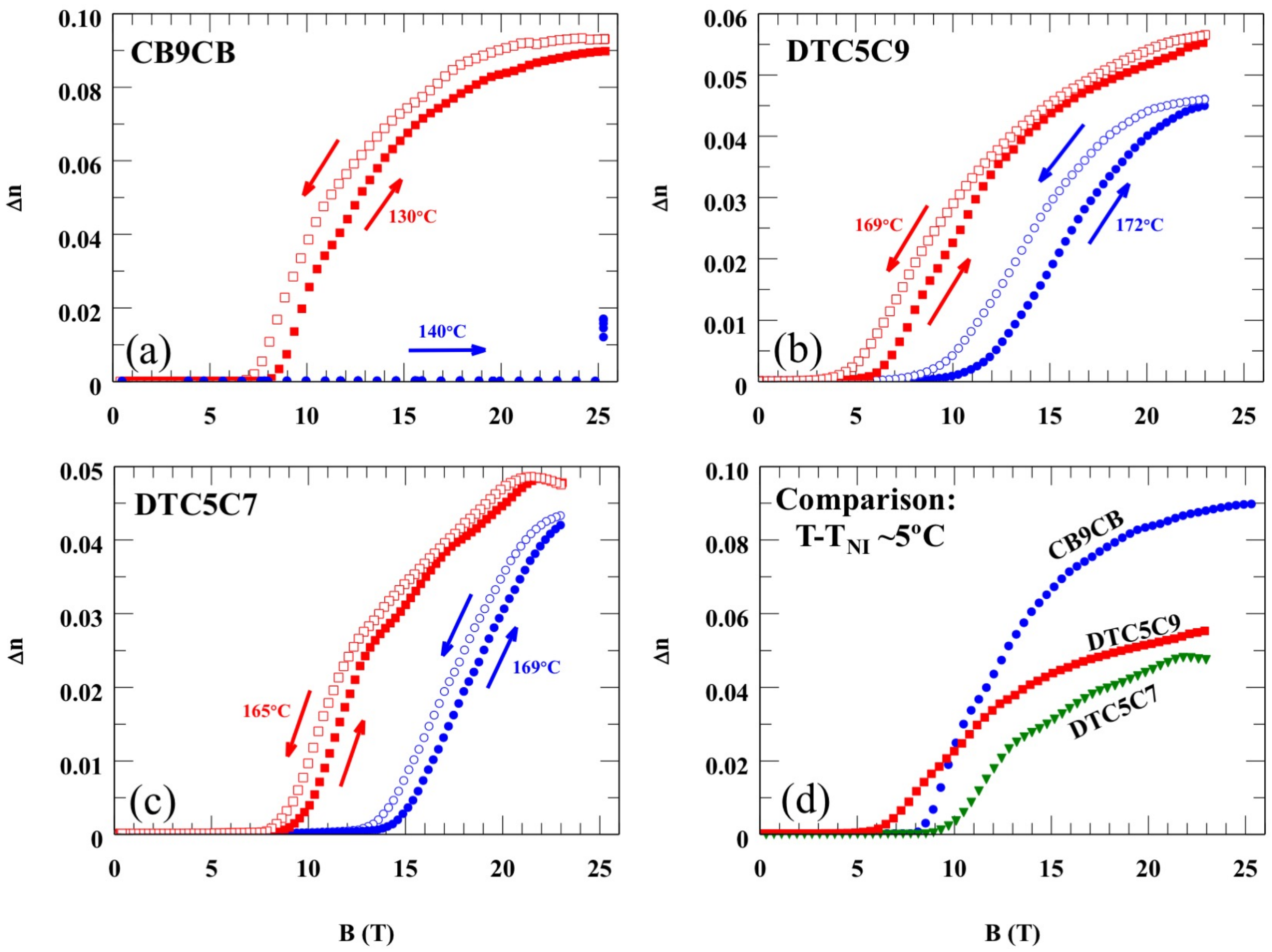

*Figure 3: Effective birefringence measurements as the function of magnetic field B. (a) CB9CB at 130ºC and 140ºC; (b) DTC5C9 at 169ºC and 172ºC; (c) DTC5C7 at 165ºC and 169ºC; The arrows ↑(↓) shows measurements at field ramping up (down) at 5ºC/min rate. (d) Comparison of the three materials at 5ºC above the zero-field $T_{N\text{-}I}$.*

Magnetic fields interact with liquid crystals mainly via the molecules' aromatic ring moieties. The free energy contribution due to an applied field will be minimized when the ring plane normals are perpendicular to the field. In fully rigid molecules this occurs with the molecular long axis aligning parallel to the field. The most profound effect of this interaction is the reorientation of rigid rod-shaped molecules towards the external field. Additionally, when the director is already aligned along the field, a magnetic field can couple to the magnitude of the orientational order, which can be described by Landau-deGennes [31] or Maier-Saupe [32] theories. In both cases, a magnetic field leads to $\Delta T_{NI}$ that is linear in $\Delta\chi_0$, the diamagnetic anisotropy and quadratic in $B$. In the Landau-deGennes theory, $\Delta T_{N\text{-}I}$ also depends inversely on the latent heat of the N-I transition. However, in order for this mechanism to explain the magnitude of $\Delta T_{N\text{-}I}$ observed, these compounds would require either i) diamagnetic anisotropy hundreds of times larger than in typical rod-shaped thermotropic compounds (which may be ruled out due to their similar cores based upon aromatic rings), or ii) N-I latent heat thousands of times smaller, which is also not the case. [33]

Hence we conclude that field-enhancement of nematic order, as predicted by the Landau-de Gennes theory for simple rodlike mesogens, cannot explain the observed values for $\Delta T_{NI.}$ The dimeric compounds show no indications of other varieties of ordering (or fluctuations thereof), such as positional clustering [17,23,25] or biaxiality, which have been proposed to explain otherwise inexplicably large $\Delta T_{NI}$. [12,13]

Most theories relating molecular morphology to the onset of the N-I transition derive from Onsager [34]; however, approaches based only on excluded volume effects are athermal and cannot provide insight to transition temperatures. Onsager-type models have been hybridized [35] [36] to contain the effects of both anisotropic, rigid shapes and anisotropic inter-molecular interactions (treated via mean-field). While these models do not aspire to describe the more complex shaped mesogens in the present work, they do indicate that $T_{NI}$ increases strongly with mesogens' aspect ratio (length/diameter). This fits broadly with experimental results for linear rigid aromatic ring systems . [35–37]

Owing to the odd numbered methylene groups in the linking group, the average molecular shape of the materials we studied is bent, (i.e. non-zero $\beta_0$ as shown in Fig 4). However, due to the inherent flexibility of the alkyl bridge between the two arms, the shape can be easily altered. A decrease of the bend angle effectively increases the aspect ratio of the dimer, which should produce a significant increase of $T_{NI}$, as recently predicted. [38] This suggests that a magnetic field-induced straightening of the molecules is responsible for the magnetically-induced large shift of $T_{NI}$ recorded in our experiments in high fields. The role of the molecular shape on $T_{NI}$ is also evidenced in the anomalously large "odd-even" effects observed in the $T_{NI}$ transition temperature. For an even number of methylene linkages (that promote a straight or linear molecular conformation) the phase transition temperatures for CBnCB molecules were found to be 50ºC higher than for the odd numbered homologs that possess a more bent shape. [39], [40]

Our hypothesis is therefore that the effect of a large field is that it “straightens” out the dimers on average, aligning the two rigid arms more parallel to a common axis than would be found in zero field. This mechanism is illustrated in Figure 4. We note, the effect of the shape, i.e., the odd-even effect, is decreasing toward higher homologs, which is corroborated by our observation (see Figure 2) that the field-induced phase sift is smaller for DTC5C9 than for DTC5C7.

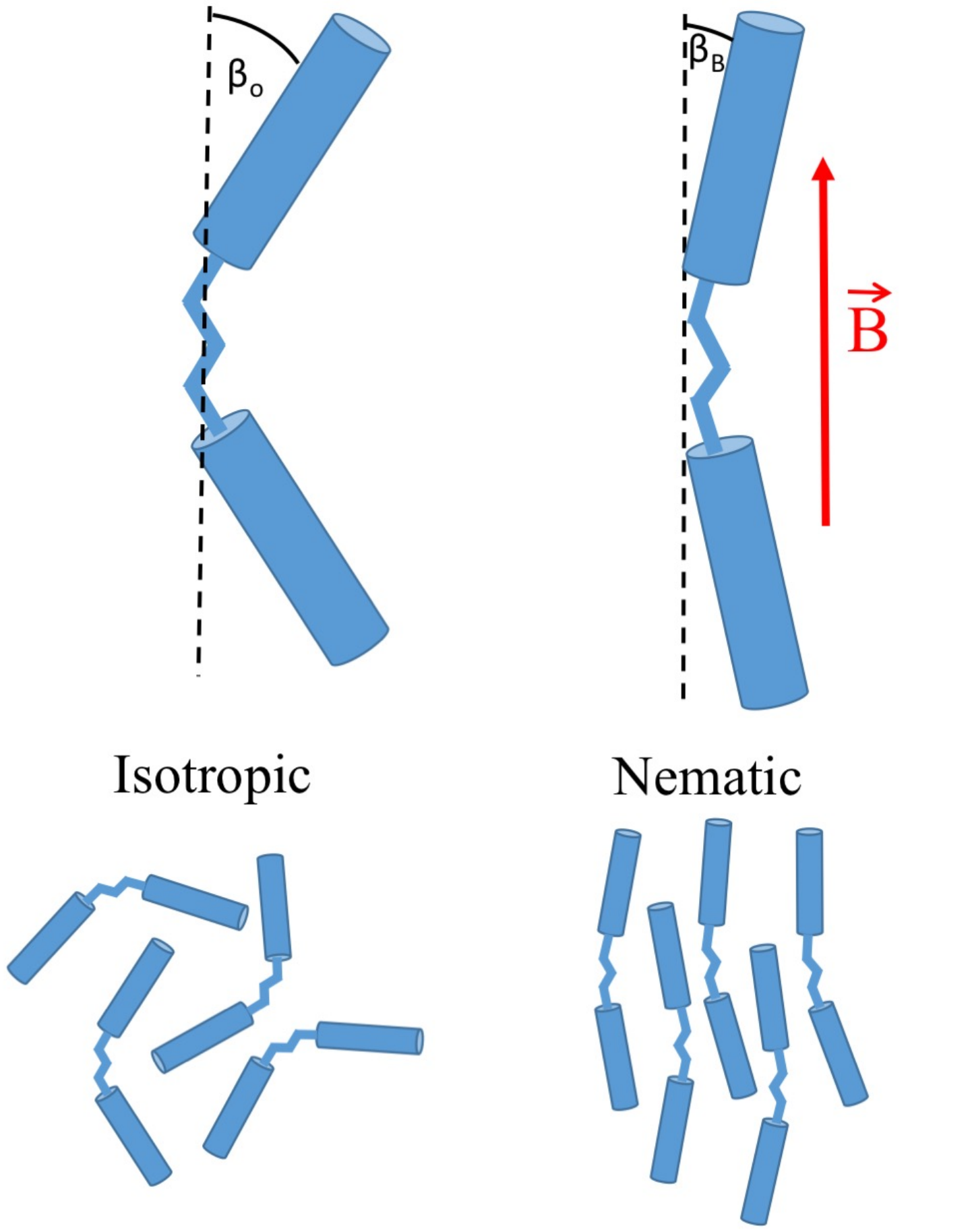

*Figure 4: Illustration of the magnetic field induced decrease of the molecular bend, and its consequence of the shift of $T_{NI}$.*

To summarize, we have observed a large magnetic field-induced elevation in the nematic to isotropic phase transition temperature in various thermotropic liquid crystal dimers containing an odd number of methylene groups in the linkage between the two terminal mesogenic moieties. We attribute this unprecedented shift to a field induced "straightening" in the average conformation of the dimers. The impact of an electromagnetic field at the molecular level on the ordering of an ensemble of mesogens has not (to our knowledge) been previously reported.

**Acknowledgement:** *This work was financially supported by NSF DMR No. 1307674 and utilized the facilities of the NHMFL, which is supported by NSF DMR-0084173, the State of Florida, and the US Department of Energy and the EPSRC (UK) through project EP/J004480/1.*